\documentclass[aps,prl,preprintnumbers,twocolumn,groupedaddress,nofootinbib]{revtex4}

\usepackage{latexsym}
\usepackage{amsmath}
\usepackage{amsfonts}
\usepackage{amsthm}
\usepackage{graphicx}

\usepackage{color}

\usepackage[latin1,utf8]{inputenc}
\usepackage[T1]{fontenc}

\allowdisplaybreaks

\newcommand{\bq}{\begin{eqnarray}}
\newcommand{\eq}{\end{eqnarray}}
\newcommand{\eps}{\varepsilon}
\newcommand{\gfour}{K}
\newcommand{\qbar}{q}

\theoremstyle{plain}

\newcommand{\arxivdate}{November 29, 2022}

\begin{document}

\preprint{MITP/22-094, TUM-HEP-1431/22}
\title{\boldmath{Taming Calabi--Yau Feynman integrals: The four-loop equal-mass banana integral}}

\author{Sebastian P\"ogel${}^{a}$, Xing Wang${}^{b}$ and Stefan Weinzierl${}^{a}$}
\affiliation{${}^{a}$ PRISMA Cluster of Excellence, Institut f{\"u}r Physik, Johannes Gutenberg-Universit\"at Mainz, D-55099 Mainz, Germany \\
${}^{b}$ Physik Department, James-Frank-Stra{\ss}e 1, Technische Universit\"at M\"unchen,
D - 85748 Garching, Germany
}

\date{\arxivdate}

\begin{abstract}
Certain Feynman integrals are associated to Calabi--Yau geometries.
We demonstrate how these integrals can be computed with the method of differential equations.
The four-loop equal-mass banana integral is the simplest Feynman integral whose geometry is a non-trivial Calabi--Yau manifold.
We show that its differential equation can be cast into an $\varepsilon$-factorised form.
This allows us to obtain the solution to any desired order in the dimensional regularisation parameter $\varepsilon$.
The method generalises to other Calabi--Yau Feynman integrals.
Our calculation also shows that the four-loop banana integral is only minimally more complicated 
than the corresponding Feynman integrals at two or three loops.
\end{abstract}

\maketitle

\section{Introduction}
\label{sect:intro}

In recent years it has been become clear that we may associate to each Feynman integral a geometric object,
and that understanding this geometry helps in computing the Feynman integral.
Most Feynman integrals that have been computed so far are associated to a sphere.
These Feynman integrals evaluate to multiple polylogarithms \cite{Goncharov_no_note,Borwein}.
The next more complicated Feynman integrals are associated to elliptic curves. These evaluate to iterated integrals of modular forms
and elliptic polylogarithms \cite{Adams:2017ejb,Broedel:2017kkb}.
These two classes of Feynman integrals (i.e. the ones associated to spheres and elliptic curves)
are by now quite well understood and in many examples we are able to transform the differential equation for these
Feynman integrals to an $\eps$-factorised form \cite{Henn:2013pwa}.
Examples for the elliptic case can be found in refs.~\cite{Adams:2018yfj,Bogner:2019lfa,Muller:2022gec,Giroux:2022wav}.
The differential equation in an $\eps$-factorised form together with values of the Feynman integrals at a boundary point is all that we need:
From this data we can easily obtain the analytic solution to any order in the dimensional regularisation parameter $\eps$.

We also know that there are Feynman integrals associated to Calabi--Yau manifolds \cite{Bourjaily:2018yfy}.
However, up to now it was not known whether the differential equation 
for a Feynman integral associated to a non-trivial Calabi--Yau manifold
can be transformed to an $\eps$-factorised form.
In this letter we show for the first time that this is indeed possible.
This implies that the Feynman integral can be solved to any order in the dimensional regularisation parameter $\eps$ (the required boundary values
are rather easily obtained).
Let us comment on what we mean by a ``non-trivial'' Calabi--Yau manifold:
Calabi--Yau $1$-folds are elliptic curves, which are well understood.
Feynman integrals associated to higher-dimensional Calabi--Yau manifolds 
can in some cases be related to elliptic curves. 
This occurs when their Picard--Fuchs operator is the symmetric product of a degree two operator of an
elliptic curve \cite{2013arXiv1304.5434B}. 
We call such Calabi--Yau manifolds ``trivial''.
The family of $l$-loop banana graphs provides examples for Feynman integrals related to Calabi--Yau $(l-1)$-folds.
For this reason it has received significant attention in recent years \cite{Groote:2005ay,Klemm:2019dbm,Bonisch:2020qmm,Bonisch:2021yfw,Kreimer:2022fxm,Forum:2022lpz}. 
The three-loop banana integral, a Calabi--Yau $2$-fold, 
is well-known to possess an operator that is a symmetric square \cite{Verrill:1996,Joyce:1972}. 
It can therefore be treated with methods similar to the elliptic case \cite{Bloch:2014qca,Primo:2017ipr,Broedel:2019kmn,Broedel:2021zij,Pogel:2022yat}. 
The first example from the family of banana
graphs that is non-trivial in the sense above is therefore the four-loop equal-mass banana integral. 
It is related to a Calabi--Yau three-fold.
This integral is for example relevant to the process
$p p \rightarrow t \bar{t}$ at $\mathrm{N}^4\mathrm{LO}$.

In this letter we show that an $\eps$-factorised differential equation for this Feynman integral exists, 
give the differential one-forms appearing in the differential equation
and show how to solve the four-loop equal-mass banana integral to all orders in the dimensional regularisation parameter $\eps$.
The benefits are three-fold: 
Firstly, we obtain analytic solutions to any order in the dimensional regularisation parameter.
Secondly, our results give us very fast numerical evaluations.
Thirdly, the differential one-forms define the symbol alphabet for this Feynman integral (see eq.~(\ref{alphabet})).
Hence, as a by-product we obtain for the first time the symbol alphabet for a Feynman integral associated to a Calabi--Yau three-fold,
extending recent work on elliptic symbols \cite{Kristensson:2021ani,Wilhelm:2022wow}.

The method we employ is not specific to this particular Feynman integral,
and extends to other Feynman integrals related to Calabi--Yau manifolds.
For this letter we have checked that the same method works for the five-loop and six-loop equal-mass banana integrals.
A few weeks after this letter appeared on the preprint server, two publications appeared
showing the application of methods similar to the one introduced here
to equal-mass banana integrals \cite{Pogel:2022vat} and ice cone integrals \cite{Duhr:2022dxb}
at any loop order.
In general,
Feynman integrals depend on more than one kinematic variable.
Establishing an $\eps$-factorised form for these will require similar steps as going from 
the equal-mass sunrise (one kinematic variable) to the unequal-mass sunrise (three kinematic variables).
The sunrise integrals are related to Calabi--Yau one-folds, 
and an $\eps$-factorised form exists for both equal and unequal cases. 
Hence, the results in this letter open the door for calculating Feynman integrals related to higher-dimensional
Calabi--Yau manifolds with more parameters.

We find that there are two new ingredients required for non-trivial Calabi--Yau manifolds.
The first one is a change of variables.
In the two-loop and three-loop equal-mass banana cases an important step is the change of variables from the dimensionless ratio $m^2/(-p^2)$ to
the modular parameter $\tau$, defined as the ratio of the two periods of the elliptic curve.
In the four-loop case there is no elliptic curve.
However the above change of variables can be viewed as the 
mirror map \cite{Candelas:1990rm,Batyrev:1993oya,Batyrev:1995ca}
for a family of Calabi--Yau manifolds,
and this generalises to the cases of interest.

The second ingredient is the following: The $\eps$-factorised form is achieved by redefining the master integrals.
One pattern which emerges is that one of the master integrals of the $l$-loop banana family in $D=2-2\eps$ space-time dimensions is given by
\bq
 I_2 
 & = &
 \frac{\eps^l}{\omega_1} I_{\underbrace{1 \dots 1}_{l+1}},
\eq
where $\omega_1$ is a specific solution of the homogeneous differential equation.
With $\qbar=\exp(2\pi i \tau)$ and $\theta=\qbar \frac{d}{d\qbar}$ the Picard--Fuchs operators for $I_2$ at two and three loops are
\bq
 \theta^2
 & \mbox{and} &
 \theta^3.
\eq
At four loops we find
\bq
\label{factorisation}
 \theta^2
 \frac{1}{K}
 \theta^2.
\eq
The appearance of the new function $\gfour$ is related to the fact that the Picard--Fuchs operator at four loops is not a symmetric product.
It also implies that at four loops the first term of $I_2$ in the $\eps$-expansion is not an Eichler integral.
In the literature on Calabi--Yau manifolds the function $\gfour$ is known as 
the ``Yukawa coupling''\footnote{In particle physics a Yukawa coupling describes the coupling of the Higgs boson to a fermion. 
This notion has been transferred to superstring models, compactified on Calabi--Yau three-folds. 
From there it diffused into the mathematical literature, where it is used in the context of Calabi--Yau operators. 
In this letter we are back to particle physics, but the term ``Yukawa coupling'' does not refer to the original particle physics meaning. 
In order to avoid confusion, we will not use any further the term ``Yukawa coupling''.} \cite{Morrison:1991cd,Almkvist:2004kj,Almkvist:2005,Yang:2008,2011arXiv1105.1136B,2013arXiv1304.5434B,2017arXiv170400164V,Candelas:2021tqt}.
In general the factorisation (or special local normal form) for a Calabi--Yau operator of degree $l$ is \cite{2013arXiv1304.5434B}
\bq
 \theta^2 \frac{1}{K_1} \theta \frac{1}{K_2} \theta \dots \theta \frac{1}{K_{l-4}} \theta \frac{1}{K_{l-3}} \theta^2
\eq
with $K_i=K_{l-2-i}$.
This factorisation translates into an ansatz for the $\eps$-factorised form (see eq.~(\ref{ansatz})).
In the following we discuss concretely the case of the four-loop equal-mass banana integral.
It serves as an example where complications due to higher-dimensional Calabi--Yau manifolds occur for the first time.
Our main results are the ansatz given in eq.~(\ref{ansatz}), 
the differential equation in $\eps$-factorised form given in eq.~(\ref{dgl_eq}) and eq.~(\ref{dgl_eq_ext}),
and
the symbol alphabet in eq.~(\ref{alphabet}).

\section{Notation}
\label{sect:notation}

We are interested in the integrals
\bq
\label{def_banana_loop_integral}
\lefteqn{
 I_{\nu_1 \nu_2 \nu_3 \nu_4 \nu_5}
 = 
 e^{4 \gamma_E \eps}
 \left(m^2\right)^{\sum\limits_{j=1}^5 \nu_j-2D}
 \int \left( \prod\limits_{a=1}^{5} \frac{d^Dk_a}{i \pi^{\frac{D}{2}}} \right)
 } & & \nonumber \\
 & &
 i \pi^{\frac{D}{2}} \delta^D\left(p-\sum\limits_{b=1}^{5} k_b \right)
 \left( \prod\limits_{c=1}^{5} \frac{1}{\left(-k_c^2+m^2\right)^{\nu_c}} \right),
\eq
where $D$ denotes the number of space-time dimensions, $\eps$ the dimensional regularisation parameter
and $\gamma_E$ the Euler-Mascheroni constant.
We consider these integrals in $D=2-2\eps$ space-time dimensions.
\begin{figure}
\begin{center}
\includegraphics[scale=0.8]{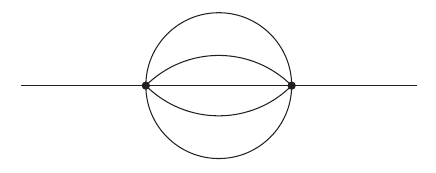}
\end{center}
\caption{
The four-loop banana graph.
}
\label{fig_banana}
\end{figure}
The corresponding Feynman graph is shown in fig.~\ref{fig_banana}.
It is convenient to introduce the dimensionless variable
\bq
 y & = & - \frac{m^2}{p^2}.
\eq
It is well-known that this family of Feynman integrals has five master integrals.
A possible choice for a basis of master integrals is
\bq
 I_{11110}, 
 \;
 I_{11111}, 
 \;
 I_{11112}, 
 \;
 I_{11113},
 \;
 I_{11114}.
\eq
The Feynman integral $I_{11110}$ is a product of four one-loop tadpole integrals
and rather simple.
We set
\bq
 I_1
 & = & 
 \eps^4 I_{11110}
 \; = \; 
 \left[ e^{\gamma_E \eps} \Gamma\left(1+\eps\right) \right]^4. 
\eq
The integral $I_1$ has uniform transcendental weight, where we assign weight $n$ to the zeta value $\zeta_n$
and weight $(-1)$ to $\eps$.

Let us now look at $I_{11111}$. Useful information can be extracted from the Picard--Fuchs operator \cite{MullerStach:2011ru,MullerStach:2012mp,Adams:2017tga,Lairez:2022zkj}.
For $I_{11111}$ we have the inhomogeneous fourth-order differential equation
\bq
 L_4 \; I_{11111}
 & = &
 \frac{120 \eps^4}{y^3\left(1+y\right)\left(1+9y\right)\left(1+25y\right)} I_{11110}.
\eq
The dependence on $\eps$ of the fourth-order differential operator $L_4$ is polynomial.
The $\eps$-independent part of $L_4$, which we denote by $L_4^{(0)}$, is of particular importance:
\bq
 L_4^{(0)}
 & = &
  \frac{d^4}{dy^4}
  + 2 \left(
           \frac{1}{y} + \frac{1}{1+y} + \frac{9}{1+9y} + \frac{25}{1+25y} 
    \right) \frac{d^3}{dy^3}
 \nonumber \\
 & &
  + \frac{\left(1+98y+1839y^2+3150y^3\right)}{y^2\left(1+y\right)\left(1+9y\right)\left(1+25y\right)} \frac{d^2}{dy^2}
 \nonumber \\
 & &
 - \frac{\left(1+15y\right)\left(1-15y-60y^2\right)}{y^3\left(1+y\right)\left(1+9y\right)\left(1+25y\right)}
   \frac{d}{dy}
 \nonumber \\
 & &
 +
 \frac{1+5y}{y^4\left(1+y\right)\left(1+9y\right)\left(1+25y\right)}.
\eq
This operator appears as entry $34$ in the table of ref.~\cite{Almkvist:2005}.
The associated Calabi--Yau manifold has been studied in ref.~\cite{Hulek:2005}.

The indicial equation for the operator $L_4^{(0)}$ at the point $y=0$ is $(\rho-1)^4=0$, 
showing that $y=0$ is a point of maximal unipotent monodromy.

\section{Master integrals}
\label{sect:master_integrals}

The construction of the master integrals, which put the differential equation into an $\eps$-factorised form,
follows with one exception the pattern found at two-loops \cite{Adams:2018yfj} and three-loops \cite{Pogel:2022yat}.
Let 
\bq
\label{lst_fct_ansatz}
 & &
 \omega_1\left(y\right),
 \;
 J\left(y\right),
 \;
 \gfour\left(y\right),
 \\
 & &
 F_{32}\left(y\right),
 \;
 F_{42}\left(y\right),
 \;
 F_{43}\left(y\right),
 \;
 F_{52}\left(y\right),
 \;
 F_{53}\left(y\right),
 \;
 F_{54}\left(y\right)
 \nonumber
\eq
be nine a priori unknown functions of $y$. 
They are however independent of $\eps$.
We start from the following ansatz for the 
master integrals
\bq
\label{ansatz}
 I_1
 & = &
 \eps^4 I_{11110},
 \nonumber \\
 I_2
 & = &
 \frac{\eps^4}{\omega_1} I_{11111},
 \nonumber \\
 I_3
 & = &
 \frac{J}{\eps} \frac{d}{dy} I_2
 + F_{32} I_2,
 \nonumber \\
 I_4
 & = &
 \frac{J}{\eps \gfour} \frac{d}{dy} I_3
 + F_{42} I_2 
 + F_{43} I_3,
 \nonumber \\
 I_5
 & = &
 \frac{J}{\eps} \frac{d}{dy} I_4
 + F_{52} I_2 
 + F_{53} I_3
 + F_{54} I_4.
\eq
The new ingredient at four loops is the appearance of the function $K$ in the definition of $I_4$.
This ansatz leads to the differential equation
\bq
 d I & = & A I,
\eq
where $A$ is a $5\times 5$-matrix.
The entries in the first four rows and the entry $A_{5,1}$ are already proportional to $\eps$.
The remaining entries in the fifth row may be written as
\bq
\label{non_eps_terms}
 A_{5,k} \; = \; \sum\limits_{j=k-5}^1 A_{5,k}^{(j)} \eps^j,
 & & k \in \{2,3,4,5\},
\eq
where $A_{5,k}^{(j)}$ is independent of $\eps$.
We now require that the $A_{5,k}^{(j)}$ with $j<1$ vanish. This gives ten equations for nine unknown functions.
These equations can be satisfied as follows:
Let $\omega_1$--$\omega_4$ be four independent solutions of $L_4^{(0)} \omega = 0$.
From Frobenius it follows that we may write them as
\bq
\label{Frobenius_series}
 \omega_k 
 & = &
 \frac{1}{\left(2\pi i\right)^{k-1}}
 \sum\limits_{j=0}^{k-1}
 \frac{\ln^j y}{j!} 
 \sum\limits_{n=0}^\infty 
 a_{k-1-j,n} y^{n+1}.
\eq
As normalisation we choose $a_{0,0}=1$.
We identify the holomorphic solution $\omega_1$ with the one appearing in eq.~(\ref{lst_fct_ansatz}) and eq.~(\ref{ansatz}).
The first few terms read
\bq
 \omega_1 & = &
 y \left(1-5y+45y^2-545y^3\right) + {\mathcal O}\left(y^5\right).
\eq
We set
\bq
\label{mirror_map}
 \tau = \frac{\omega_2}{\omega_1},
 & &
 \qbar = e^{2\pi i \tau}
\eq
and
\bq
 J & = & \frac{1}{2\pi i} \frac{dy}{d\tau}.
\eq
Eq.~(\ref{mirror_map}) defines a change of variables from $y$ to $\tau$ (or $\qbar$). 
This map is also called the mirror map\footnote{In mathematics the mirror map connects a Calabi--Yau $n$-fold with Hodge numbers $h^{p,q}$ to another Calabi--Yau $n$-fold with Hodge numbers $h^{n-p,q}$.}.
We have
\bq
 y
 & = & 
 \qbar + 8 \qbar^2 + 36 \qbar^3 + 168 \qbar^4
 + {\mathcal O}\left(\qbar^5\right).
\eq
As $y$ goes to zero, $\tau$ approaches $i\infty$ and $\qbar$ goes to zero.
The function $\gfour$ is given by
\bq
 \gfour
 & = & 
 \frac{J^3}{\omega_1^2} \frac{1}{y\left(1+y\right)\left(1+9y\right)\left(1+25y\right)}.
\eq
We define iterated integrals by
\bq
\lefteqn{
I\left(f_1,...,f_n;\tau\right)
 = } & & \\
 & &
 \lim\limits_{\qbar_0\rightarrow 0}
 R \left[
 \int\limits_{\qbar_0}^{\qbar} \frac{d\qbar_1}{\qbar_1}
 ...
 \int\limits_{\qbar_0}^{\qbar_{n-1}} \frac{d\qbar_n}{\qbar_n}
 \;
 f_1\left(\tau_1\right)
 ...
 f_n\left(\tau_n\right)
 \right],
 \nonumber
\eq
where $R$ is the operator which removes all $\ln(\qbar_0)$-terms, 
corresponding to the standard ``trailing zero'' or ``tangential base point''
regularisation \cite{Brown:2014aa,Adams:2017ejb,Walden:2020odh}.
As the last argument of all iterated integrals will always be $\tau$, we simply write $I(f_1,\dots,f_n)$ instead of $I(f_1,\dots,f_n;\tau)$.  
The functions $F_{ij}$ are given by
\begin{align}
 F_{32} & = - f_{2,a} - f_{2,b},
 &
 \\
 F_{43} & = - \frac{1}{K} \left( f_{2,a} - f_{2,b} \right),
 &
 F_{42} & = -\frac{1}{K} f_{4,a},
 \nonumber \\
 F_{54} & =  - f_{2,a}+ f_{2,b},
 &
 F_{53} & = - f_{4,b},
 &
 F_{52} & = - f_6,
 \nonumber
\end{align}
where
\bq
 f_{2,a}
 & = & 
 J \left( \frac{5}{2y} - \frac{1}{1+y} - \frac{9}{1+9y} - \frac{25}{1+25y} \right),
 \\
 f_{2,b}
 & = & 
 I\left(K,h_6\right),
 \nonumber \\
 f_{4,a}
 & = & 
 - \left[ I\left(K,h_6\right) \right]^2
 - 2 K I\left(1,K,h_6,h_6\right)
 \nonumber \\
 & &
 - K I\left(1,h_{8,b}\right),
 \nonumber \\
 f_{4,b}
 & = &
 4 I\left(1,K,h_6,h_6\right)
 + 2 I\left(1,h_{8,b}\right)
 + h_4,
 \nonumber \\
 f_5
 & = &
 120 \frac{J \omega}{y^2},
 \nonumber \\
 f_6
 & = &
 f_{4,b} I\left(K,h_6\right) - 4 h_6,
 \nonumber \\
 f_8
 & = &
 f_{4,b} \left[ I\left(K,h_6\right) \right]^2 - 8 h_6 I\left(K,h_6\right)
 \nonumber \\
 & &
 + K \left[ I\left(1,h_{8,b}\right) + 2 I\left(1,K,h_6,h_6\right) \right]^2
 + h_{8,a}.
 \nonumber
\eq
The helper functions $h_4$, $h_6$, $h_{8,a}$ and $h_{8,b}$ are defined in the appendix.
This completes the definition of the master integrals and it can be verified that with this definition 
the terms $A_{5,k}^{(j)}$ with $j<1$ vanish in eq.~(\ref{non_eps_terms}).
Hence, the differential equation is in $\eps$-factorised form
\bq
\label{dgl_eq}
 d I
 & = &
 2 \pi i \eps A_\tau I d\tau,
\eq
with
\bq
\label{dgl_eq_ext}
\lefteqn{
 A_\tau
 = } & &
 \\
 & &
 \left( \begin{array}{ccccc}
 0 & 0 & 0 & 0 & 0 \\
 0 & f_{2,a}+f_{2,b} & 1 & 0 & 0 \\
 0 & f_{4,a} & f_{2,a}-f_{2,b} & K & 0 \\
 0 & f_6 & f_{4,b} & f_{2,a}-f_{2,b} & 1 \\
 f_5 & f_8 & f_6 & f_{4,a} & f_{2,a}+f_{2,b} \\
 \end{array} \right).
 \nonumber
\eq
This is a differential equation with an alphabet consisting of nine letters
\bq
\label{alphabet}
 {\mathcal A}
 & = &
 \left\{
  1, K, f_{2,a}, f_{2,b}, f_{4,a}, f_{4,b}, f_5, f_6, f_8
 \right\}.
\eq
We observe the symmetry
\bq
 A_{i,j} \; = \; A_{7-j,7-i}
 & &
 \mbox{for} \; i,j \ge 2.
\eq
We further observe that the $\qbar$-expansions of all entries of the matrix $A_\tau$ 
have integer coefficients (apart from a possible rational prefactor):
\bq
\label{q_expansions}
 K 
 & = & 
 1 - \qbar + 17 \qbar^2 - 253 \qbar^3 + 3345 \qbar^4 
 + {\mathcal O}\left(\qbar^5\right),
 \nonumber \\
 f_{2,a}
 & = &
 \frac{5}{2} - 15 \qbar + 167 \qbar^2 - 2787 \qbar^3 + 40631 \qbar^4
 + {\mathcal O}\left(\qbar^5\right),
 \nonumber \\
 f_{2,b}
 & = &
 7 \qbar - 135 \qbar^2 + 2275 \qbar^3 - 34759 \qbar^4
 + {\mathcal O}\left(\qbar^5\right),
 \nonumber \\
 f_{4,a}
 & = &
 - \frac{1}{4} \left( 147 \qbar -3267 \qbar^2 + 59943 \qbar^3 - 1017027 \qbar^4 \right)
 \nonumber \\
 & &
 + {\mathcal O}\left(\qbar^5\right),
 \nonumber \\
 f_{4,b}
 & = &
 \frac{5}{2} - 52 \qbar + 1460 \qbar^2 - 33316 \qbar^3 + 652212 \qbar^4
 \nonumber \\
 & &
 + {\mathcal O}\left(\qbar^5\right),
 \nonumber \\
 f_{5}
 & = &
 120 \left(1 + 3 \qbar - 27 \qbar^2 + 147 \qbar^3 - 1467 \qbar^4 \right)
 + {\mathcal O}\left(\qbar^5\right),
 \nonumber \\
 f_{6}
 & = &
 - \frac{1}{2} \left( 21 \qbar - 2805 \qbar^2 + 108777 \qbar^3 - 2772213 \qbar^4 \right)
 \nonumber \\
 & &
 + {\mathcal O}\left(\qbar^5\right),
 \nonumber \\
 f_{8}
 & = &
 - \frac{1}{16} \left( 9 - 855 \qbar + 7623 \qbar^2 + 606789 \qbar^3
 \right. \nonumber \\
 & & \left.
                       - 31766841 \qbar^4 \right)
 + {\mathcal O}\left(\qbar^5\right).
\eq

\section{Results}
\label{sect:results}

We write
\bq
 I_k
 & = &
 \sum\limits_{j=0}^\infty I_k^{(j)} \eps^j
\eq
for the $\eps$-expansion of the master integral $I_k$.

With the differential equation~(\ref{dgl_eq}) at hand we only need the boundary values as additional input.
We choose $y=0$ as boundary point.
The boundary values are easily obtained with the help of the Mellin--Barnes technique.
The calculation follows the lines of the corresponding calculation at three loops \cite{Broedel:2019kmn,Pogel:2022yat}.
We need the constant term and all logarithms $\ln(y)$.
We obtain for the boundary value of $I_2$ \cite{Bonisch:2021yfw}
\bq
\label{boundary_value}
 \left. I_2 \right|_{y \to 0}
 & = &
 5  e^{4 \gamma_E \eps}
 \sum\limits_{j=0}^4
  \left( \begin{array}{c} 4 \\ j \\ \end{array}\right)
  \left(-1\right)^j
  y^{j\eps}
 \\
 & &
  \frac{\Gamma\left(1+\eps\right)^{4-j}\Gamma\left(1-\eps\right)^{1+j}\Gamma\left(1+j\eps\right)}{\Gamma\left(1-\left(j+1\right)\eps\right)}.
 \nonumber
\eq
Note that eq.~(\ref{boundary_value}) determines the boundary values of all master integrals.

With the boundary values we may now obtain analytical results for all master integrals. 
As an example we focus on $I_2$.
With
\bq
 f_2^+ = f_{2,a}+f_{2,b},
 & &
 f_2^- = f_{2,a}-f_{2,b}
\eq
the analytic result for the integral $I_2$ up to ${\mathcal O}(\eps^5)$ is given 
by
\bq
\lefteqn{
 I_2
 = 
 \eps^4 \left[ I\left(1,K,1,f_5\right) - 80 \zeta_3 \ln\qbar \right]
 } & &
 \\
 & &
 + \eps^5
 \left\{
  I\left(1,K,1,f_2^+,f_5\right)
  + I\left(1,K,f_2^-,1,f_5\right)
 \right. \nonumber \\
 & & \left.
  + I\left(1,f_2^-,K,1,f_5\right)
  + I\left(f_2^+,1,K,1,f_5\right)
 \right. \nonumber \\
 & & \left.
  - 120 \zeta_2 I\left(1,K,1\right)
  - 40 \zeta_3 \left[ 13 I\left(1,K\right)
                      + 2 I\left(1,f_2^-\right)
 \right. \right. \nonumber \\
 & & \left. \left.
                      + 2 I\left(f_2^+,1\right) \right]
  - 120 \zeta_4 \ln\qbar
  + 80 \zeta_2\zeta_3
  - 600 \zeta_5
 \right\}
 + {\mathcal O}\left(\eps^6\right).
 \nonumber
\eq
From eq.~(\ref{q_expansions}) we obtain the $\qbar$-expansion of all iterated integrals and hence the $\qbar$-expansion
of the master integrals.
For example, the first few terms of $I_2^{(4)}$ read with $L=\ln \qbar$
\bq
 I_2^{(4)}
 & = &
 5 L^4 - 80 \zeta_3 L
 + 60 L \left( 4 - L \right) \qbar
 \\
 & &
 + 15 \left(6-34L+17L^2\right) \qbar^2
 \nonumber \\
 & &
 - \frac{10}{9} \left( 123 -2024L +1518 L^2\right) \qbar^3
 \nonumber \\
 & &
 - \frac{25}{8} \left(851 +4014 L - 4014 L^2 \right) \qbar^4
 + {\mathcal O}\left(\qbar^5\right).
 \nonumber
\eq
In addition we must compute the value of $\qbar$ from a given value $y$.
In the region $|y|<1/25$ we may use the series for $\omega_1$ and $\omega_2$ given in eq.~(\ref{Frobenius_series})
to obtain from a given value of $y$ the corresponding value of $\qbar$.
Note that this is different from the situation at two and three loops, 
where we may use complete elliptic integrals to obtain the value of $\qbar$ from $y$.

\begin{figure}[t]
\begin{center}
\includegraphics[scale=0.5]{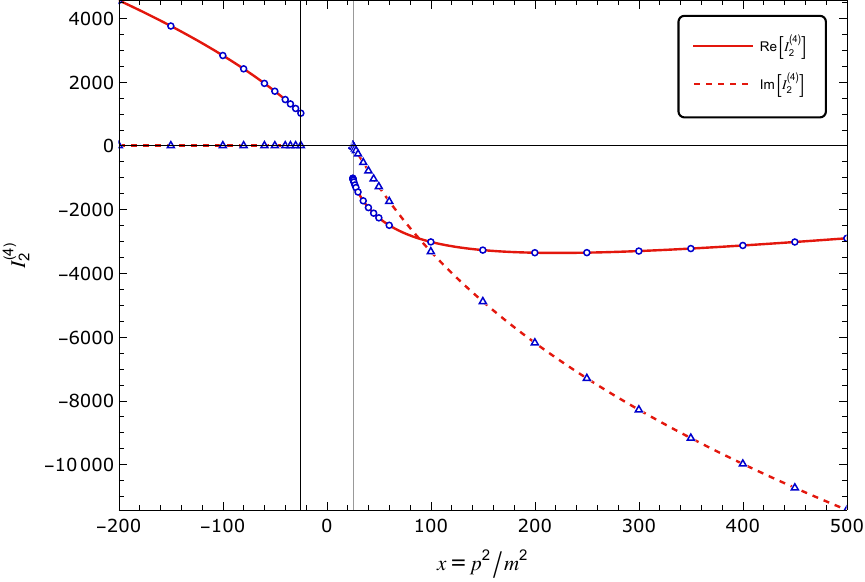}
\end{center}
\caption{
Comparison of our result for $I_2^{(4)}$ (plotted as curves) with numerical results from {\tt{pySecDec}} (points).
}
\label{fig_result_eps_4}
\end{figure}

\begin{figure}[t]
\begin{center}
\vspace*{5mm}
\includegraphics[scale=0.5]{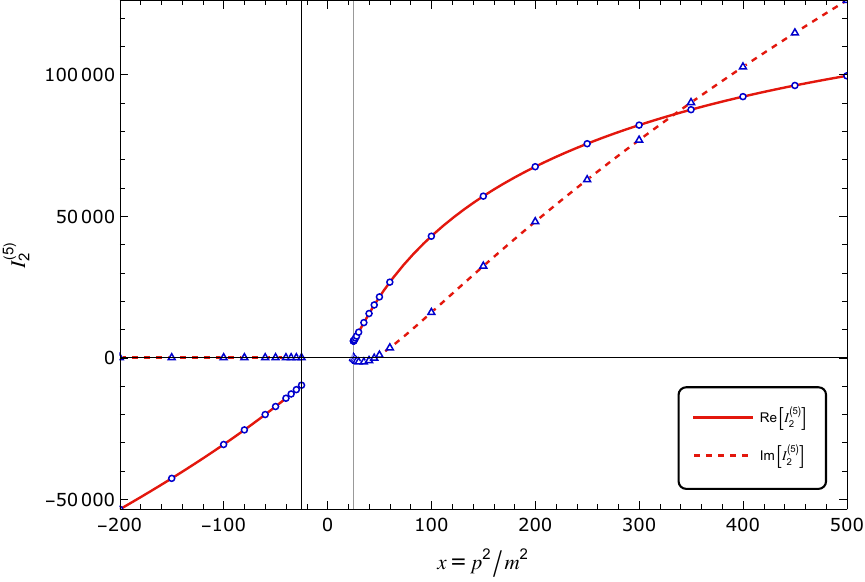}
\end{center}
\caption{
Comparison of our result for $I_2^{(5)}$ (plotted as curves) with numerical results from {\tt{pySecDec}} (points).
}
\label{fig_result_eps_5}
\end{figure}

In the plots we will use the kinematical variable $x=-1/y=p^2/m^2$ instead of $y$.
The condition $|y|<1/25$ translates to $|x|>25$. This is the region where we may evaluate the integral numerically.
The correct branch is selected by giving $x$ an infinitesimal positive imaginary part according to Feynman's $i\delta$-prescription. 
Fig.~\ref{fig_result_eps_4} and fig.~\ref{fig_result_eps_5} show the numerical results for 
the $\eps^4$-term $I_2^{(4)}$ and the $\eps^5$-term $I_2^{(5)}$ of $I_2$.
We also plotted the results from the program $\verb|pySecDec|$ \cite{Borowka:2017idc}.
We observe excellent agreement.
The CPU-time for the numerical evaluation of our result is negligible, 
a kinematic point by $\verb|pySecDec|$ takes about $10 \; \mathrm{min}$ on a desktop.

Numerical evaluations around the singular points $x=0$, $x=1$, $x=9$ and $x=25$ 
will be discussed in a longer publication.

\section{Conclusions}
\label{sect:conclusions}

In this letter we showed that the differential equation for the four-loop equal-mass banana graph
can be cast into an $\eps$-factorised form, which allows for a systematic solution to any desired order
in the dimensional regularisation parameter.

We used the mirror map to define new variables $\tau$ and $\qbar$.
The relevant Picard--Fuchs operator is no longer a symmetric product. 
We showed that this is not an essential complication, it merely introduces the function $K$.

Our calculation shows that the four-loop equal-mass banana integral
is only minimally more complicated 
than the corresponding Feynman integrals at two or three loops.
This is good news, as it opens a path towards Feynman integrals
related to genuinely generic Calabi--Yau manifolds.

\subsection*{Acknowledgements}

We would like to thank Duco van Straten for useful discussions.
This work has been supported
by the Cluster of Excellence Precision Physics, Fundamental Interactions, and Structure of
Matter (PRISMA EXC 2118/1) funded by the German Research Foundation (DFG) within
the German Excellence Strategy (Project ID 39083149).


\begin{appendix}

\section{Appendix: Helper functions}

In this appendix we give the explicit expressions for the helper functions $h_4$, $h_6$, $h_{8,a}$ and $h_{8,b}$:
\begin{widetext}
\bq
 h_4
 & = &
 \frac{5+14y-569y^{2}-12044y^{3}+77427y^{4}-302850y^{5}+50625y^{6}}{2y\left(1+y\right)\left(1+9y\right)\left(1+25y\right)} \frac{\omega_1^2}{J},
 \nonumber \\
 h_6 & = &
 \left[ \frac{7}{y} 
        + \frac{192y}{\left(1+y\right)^2} 
        + \frac{64\left(1-9y\right)}{\left(1+9y\right)^2} 
        - \frac{192\left(3+25y\right)}{\left(1+25y\right)^2}
 \right] \omega_1^2,
 \nonumber \\
 h_{8,a}
 & = & 
 - \left(1+33y-577y^{2}-225y^{3}\right)\left(3-25y-187y^{2}+225y^{3}\right)
 \nonumber \\
 & &
 \frac{\left(3-38y-2167y^{2}+4060y^{3}+307893y^{4}+693450y^{5}-50625y^{6}\right)}{16y^3\left(1+y\right)^3\left(1+9y\right)^3\left(1+25y\right)^3}  \omega_1^2 J,
 \nonumber \\
 h_{8,b} & = &
 \left[
 - \frac{h_4 J}{\omega_1^2} \frac{d^2 \omega_1}{dy^2}
 - \left(\frac{d}{dy} \frac{h_4 J}{\omega_1^2} \right) \left(\frac{d\omega_1}{dy}\right)
 - \frac{X}{4y^3\left(1+y\right)^3\left(1+9y\right)^3\left(1+25y\right)^3} \omega_1
 \right] \omega_1 J,
 \nonumber \\
 X
 & = &
 10+953y+36365y^{2}+624143y^{3}+5506553y^{4}+25045562y^{5}-43328414y^{6}-1595702658y^{7}-3663006612y^{8}
 \nonumber \\
 & &
 +4836275325y^{9}+29858270625y^{10}+9146671875y^{11}+2562890625y^{12}.
\eq
\end{widetext}

\end{appendix}

\bibliography{/home/stefanw/notes/biblio}
\bibliographystyle{/home/stefanw/latex-style/h-physrev5}

\end{document}